\documentclass{webofc}
\usepackage[varg]{txfonts}   
\usepackage{lineno}

\begin{document}
\title{{Recent results on hard probes with ALICE and STAR}}
%
%

\author{\firstname{Yaxian} \lastname{Mao}\inst{1}\fnsep\thanks{\email{yaxian.mao@mail.ccnu.edu.cn}\\Supported by National Natural Science Foundation of China (11505072).}
}

\institute{ Key Laboratory of Quark \& Lepton Physics of Ministry of Education, \\
           Institute of Particle Physics, Central China Normal University, Wuhan 430079, China
          }

\abstract{%
This paper reviews recent experimental results on hard probes in heavy-ion collisions from the ALICE and STAR Collaboration. These studies include various observables characterizing jet properties like nuclear modification factors, recoil jet yields, di-jet and photon-jet energy imbalance, and the  observables characterizing jet properties like jet fragmentation function and jet shapes; and measurements of high-$p_T$ charged hadrons from jet fragmentation and triggered particle correlations will be highlighted. }
\maketitle
\section{Introduction}
\label{intro}
The experimental heavy-ion program aims to explore the phases of nuclear matter, the structure of the nucleus and the nucleons, and the properties of the Quark Gluon Plasma (QGP), the primordial state of matter that exists during a few hundreds of microseconds in the early universe until the hadron phase transition. Hot and dense de-confined matter, the QGP state, is formed by colliding heavy ions at ultra-relativistic energies~\cite{Bjorken}. The study of the QGP in the laboratory will allow us to gain a more profound understanding of quantum chromodynamics (QCD), the theory of the strong interaction. Extending the study of QCD to dynamical, complex systems will let us learn how macroscopic properties of matter emerge from the fundamental microscopic laws of particle physics. The determinations of fundamental properties of this state of matter, such as the critical temperature, the degrees of freedom, and in general the transport coefficients is the objective of heavy-ion physics. 

Transport properties, linked to the strong interaction properties at high temperature and high color-charge density as well as macroscopic properties of the QGP, can be inferred by studying how elementary particles of the Standard Model propagate through the medium. A large variety of observables are available to the experimentalist to infer the medium properties and to understand the dynamic evolution of the heavy ion collisions. On one hand, soft probes, which have transverse momentum compatible with the temperature of the system, allow us to explore the fluid dynamics and thermodynamics of the QGP and measure the relevant associated parameters. On the other hand, hard probes (high $p_T$ or high mass), that involve scales incompatible with the thermal medium, are exploited as "external" probes of the QGP to understand its interactions depending on the color charge, mass and flavor of the probe. In order to establish the (thermo)dynamics properties of the QGP itself, detailed and comprehensive measurements employing different probes have been performed at RHIC and at LHC and a wealth of information has been collected already~\cite{RHIC_RAA, CMS_Raa2012, ATLAS, ALICE}. There is a good qualitative understanding of the hard probe interactions with the medium, but the complete quantitative understanding of the heavy-ion collision dynamics is still missing.

Jet quenching is a phenomenon that has been observed for all hadron species at high $p_T$ as a pronounced suppression of the production in heavy-ion collisions relative to pp collisions. These high $p_T$ hadrons originate from hard scattered partons produced in the initial stage of the collisions and lose energy while traversing the medium. The partonic shower generated by the hard scattering process following its fragmentation is measured as a collimated jet of hadrons. Due to the jet quenching, both the jet energy and its composition can be modified by the medium. Comparing the jet fragmentation and structure measured in proton-proton and heavy-ion collisions will reveal experimentally these modifications. Studying the jet fragmentation pattern and jet structure as a function of parton color charge and quark mass can provide discriminative and crucial information on the parton-shower and energy loss mechanism that in turn will enable to characterize qualitatively and quantitatively the parton-medium interactions. It is well established that the parton energy loss is a consequence of the modification of the hard-scattering processes  in the final state of the heavy ion collisions at RHIC and LHC~\cite{RHIC_RAA,CMS_Raa2012,ATLAS,ALICE}, which is accompanied by an enhancement of low $p_T$ hadrons, suggesting a softening of jet fragmentations.  


In the following, we summarize the charged hardons $R_{AA}$, inclusive and $D^{0}$-tagged jet production and nuclear modification factor $R_{AA}$, the recoil jet yield and jet broadening using two particle correlations. The measurements of tagged jet fragmentation properties using triggered particle correlations will be also highlighted. 

\section{Nuclear modification factor $R_{AA}$}
\label{Raa}

The nuclear modification factor, $R_{\rm AA}$,  is defined as the ratio of the particle production in Pb–Pb divided by the spectrum in pp collisions scaled by the average number of binary nucleon–nucleon collisions $N_coll$:
\begin{equation}
\begin{split}
R_{\rm AA}&=\frac{1}{<N_{\rm coll}>}\frac{{\rm d}^{2}N^{\rm AA}/{\rm d} p_{\rm T}{\rm d} \eta}{{\rm d}^{2}N^{\rm pp}/{\rm d} p_{\rm T}{\rm d}\eta},
\end{split}
\end{equation}
It is constructed such that $R_{\rm AA}$ equals unity if there is no net nuclear modification of the spectrum in Pb–Pb collisions as compared to an incoherent superposition of independent pp collisions. In a simplified picture, $R_{\rm AA}$ ratio involves both cold nuclear matter effects arising from nuclear parton distribution function modifications and the parton energy loss experienced by hard probes when traversing the hot QCD medium. 

\begin{figure*}[htbp]
 \begin{center}
 \includegraphics[width=1.0\textwidth]{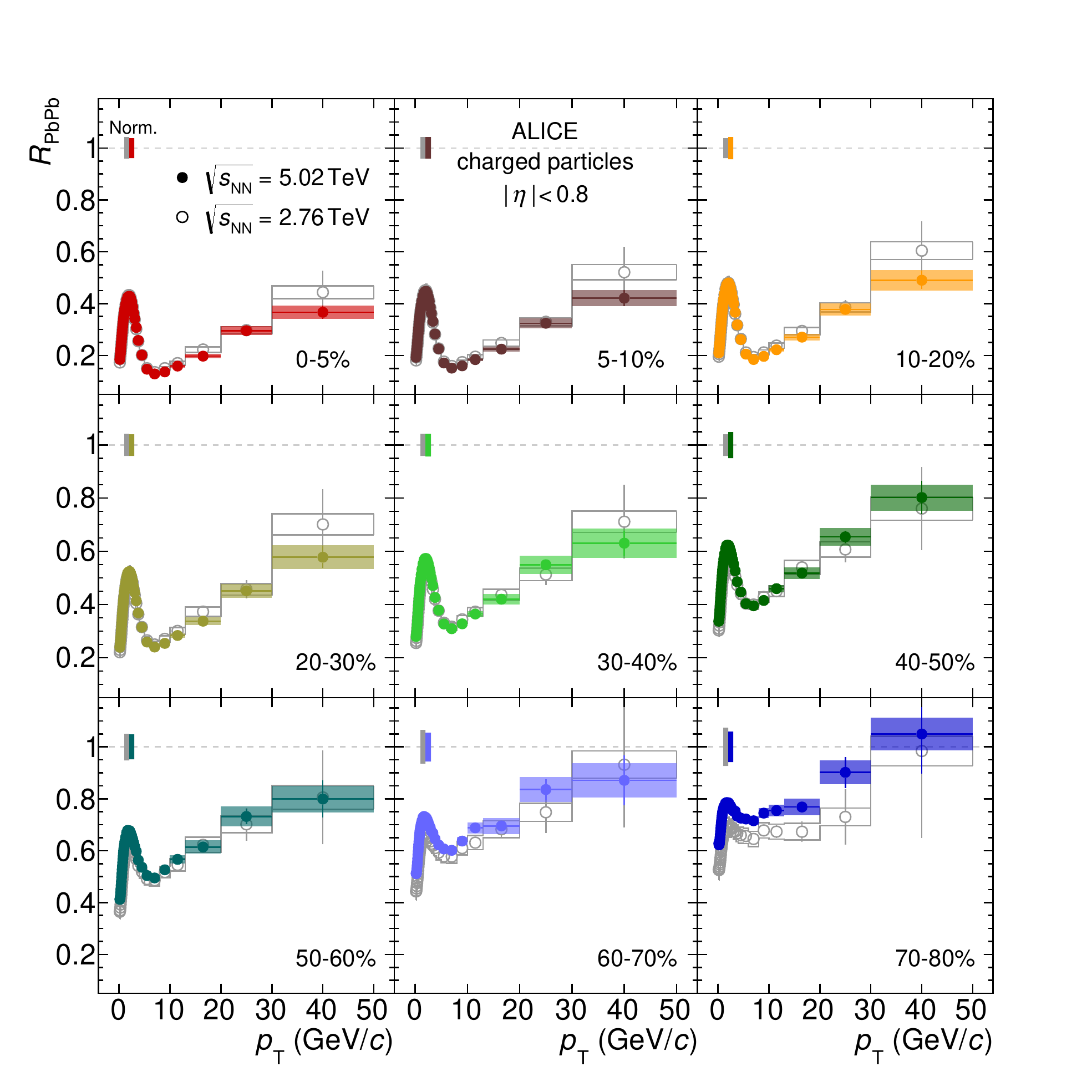}
\end{center}
\caption{Nuclear modification factors, $R_{AA}$ for inclusive charged particles in Pb-Pb collisions at $\sqrt{s_{NN}} = 2.76$ and $5.02$ TeV in different centrality intervals~\cite{hadronRAA}.}
\label{fig:hadronRaa}
\end{figure*}

Fig.~\ref{fig:hadronRaa} shows the $R_{\rm AA}$ of charged hadrons measured by ALICE in Pb-Pb collisions at $\sqrt{s_{NN}} = 2.76 $ and 5.02 TeV in different centrality bins. The nuclear modification factor has a strong centrality dependence, and is very similar in magnitude for the two collision energies. Given that the $p_T$ spectrum is harder at the higher energy, this similarity of the $R_{\rm AA}$ indicates a larger parton energy loss in the hotter/denser and longer-lived deconfined medium produced at the higher center-of-mass energy.

\begin{figure*}[htbp]
 \begin{center}
 \includegraphics[width=1.0\textwidth]{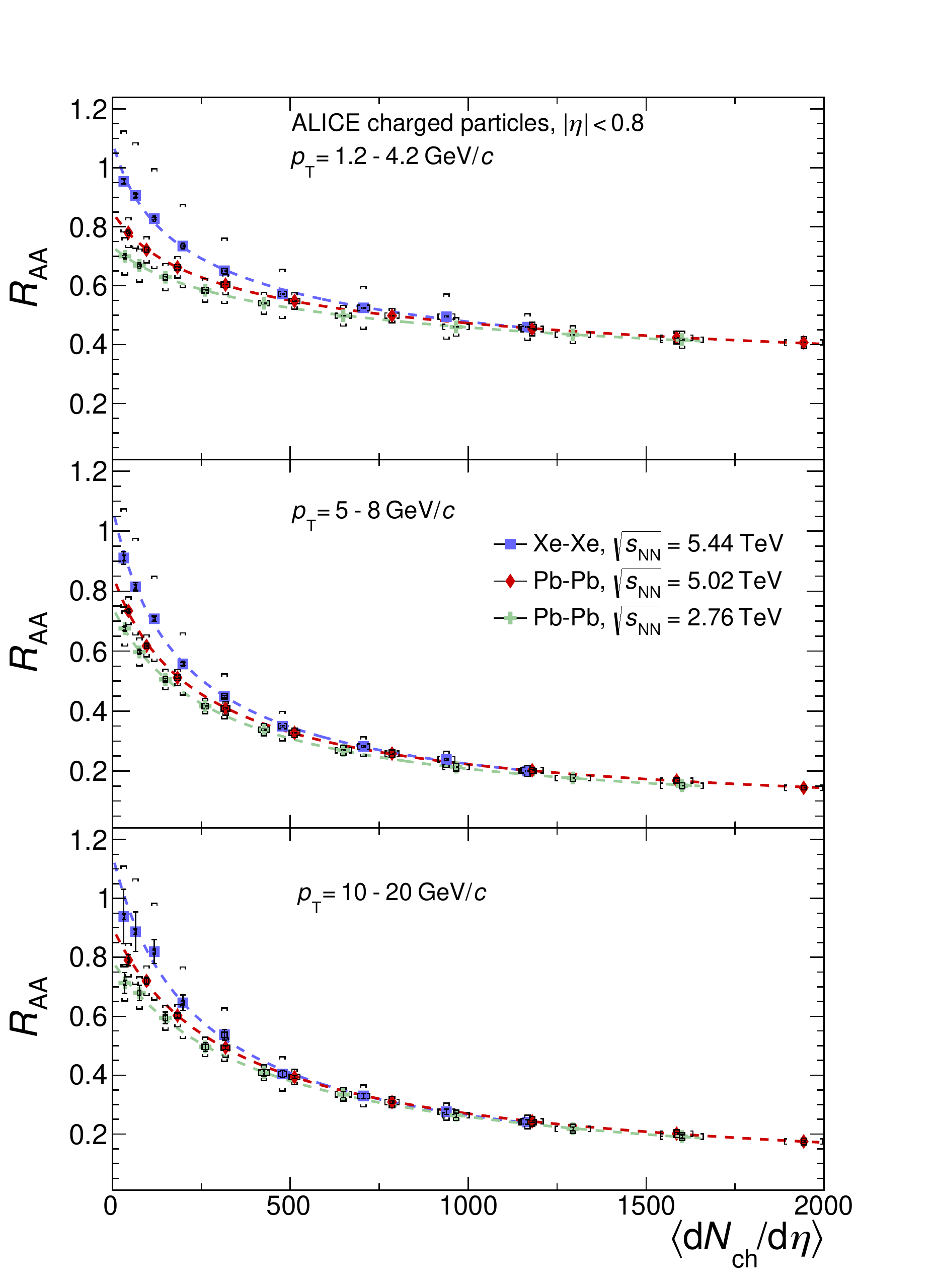}
\end{center}
\caption{Comparison of the nuclear modification factor in Xe–Xe and Pb–Pb collisions integrated over identical regions in $p_T$ as a function of $<dN_{ch}/d\eta>$~\cite{XeRAA}.}
\label{fig:RaaM}
\end{figure*}

A comparison of the nuclear modification factors as a function of $<dN_{ch}/d\eta>$ in Xe–Xe and Pb– Pb collisions for three different regions of $p_T$ (low, medium, and high) is shown in Fig.~\ref{fig:RaaM}. A remarkable similarity in $R_{AA}$ is observed between Xe–Xe collision at $\sqrt{s_{NN}} = 5.44 $ TeV and Pb–Pb collisions at $\sqrt{s_{NN}} = 5.02 $ and 2.76 TeV when compared at identical ranges in $<dN_{ch}/d\eta>$. At low $<dN_{ch}/d\eta>$, there is deviation for different collision systems, however, the values of $R_{AA}$ still agree within rather large uncertainties~\cite{XeRAA}.  

\begin{figure*}[htbp]
 \begin{center}
 \includegraphics[width=1.0\textwidth]{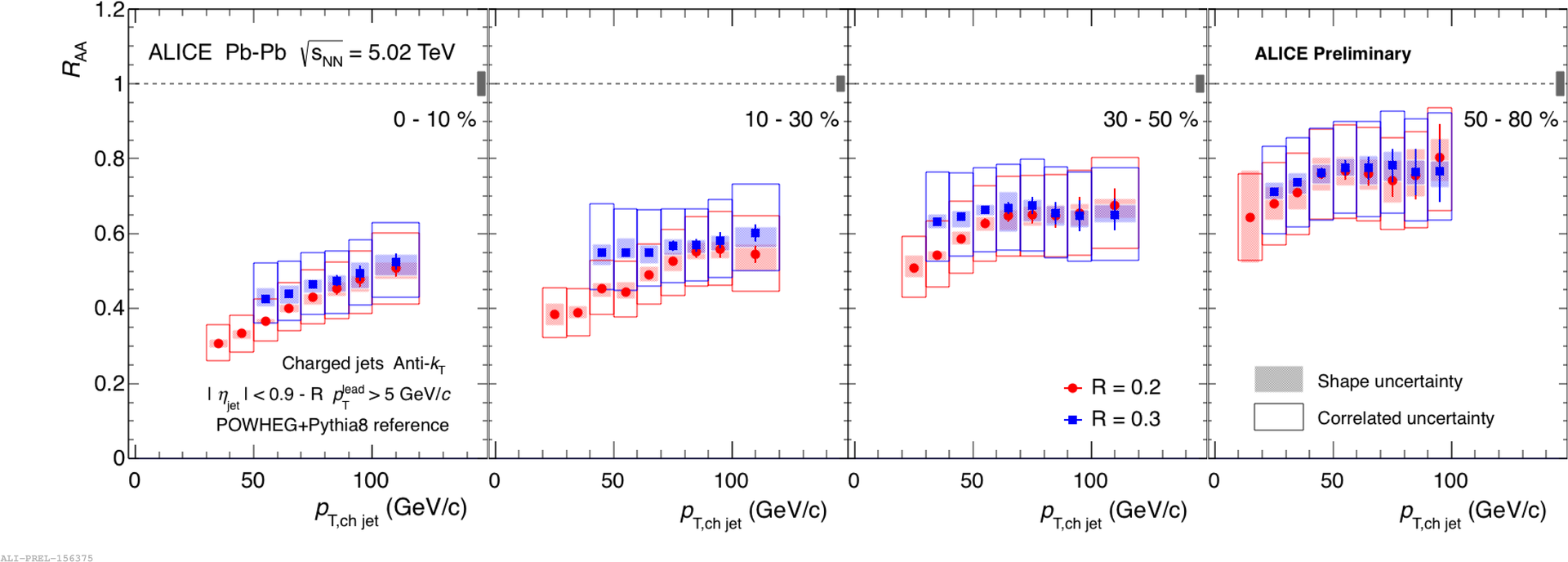}
\end{center}
\caption{Nuclear modification factors, $R_{AA}$ for charged jets with resolution parameters $R = 0.2$ and $R = 0.3$ in Pb-Pb collisions at $\sqrt{s_{NN}} = 5.02$ TeV in different centraility intervals.}
\label{fig:JetRaa}
\end{figure*}

Jets are reconstructed using the anti-$k_{\rm T}$ jet algorithm, with resolution parameters $R = 0.2$ and $R = 0.3$ by combining tracking information from ALICE detectors.
In order to measure the inclusive jet $p_T$ spectra in Pb-Pb and compare them to the production in pp collisions, and form their ratios $R_{AA}$, for each centrality bin, careful treatment of the jet energy resolutions, as well as the application of unfolding methods are necessary. 
The preliminary results of the jet nuclear modification factors between 10 and 150 GeV/$c$ of jet $p_T$ are presented in Fig.~\ref{fig:JetRaa}, in six centrality bins, 0-5\% representing the 5\% most central Pb-Pb collisions. A suppression of the jet yield reconstructed from Pb-Pb data is observed, which is increasing with jet $p_T$, and the $R_{\rm AA}$ measured by different jet resolution parameters are consistent within uncertainties. 
\label{hRAA}

\begin{figure*}[htbp]
 \begin{center}
\includegraphics[width=0.41\textwidth]{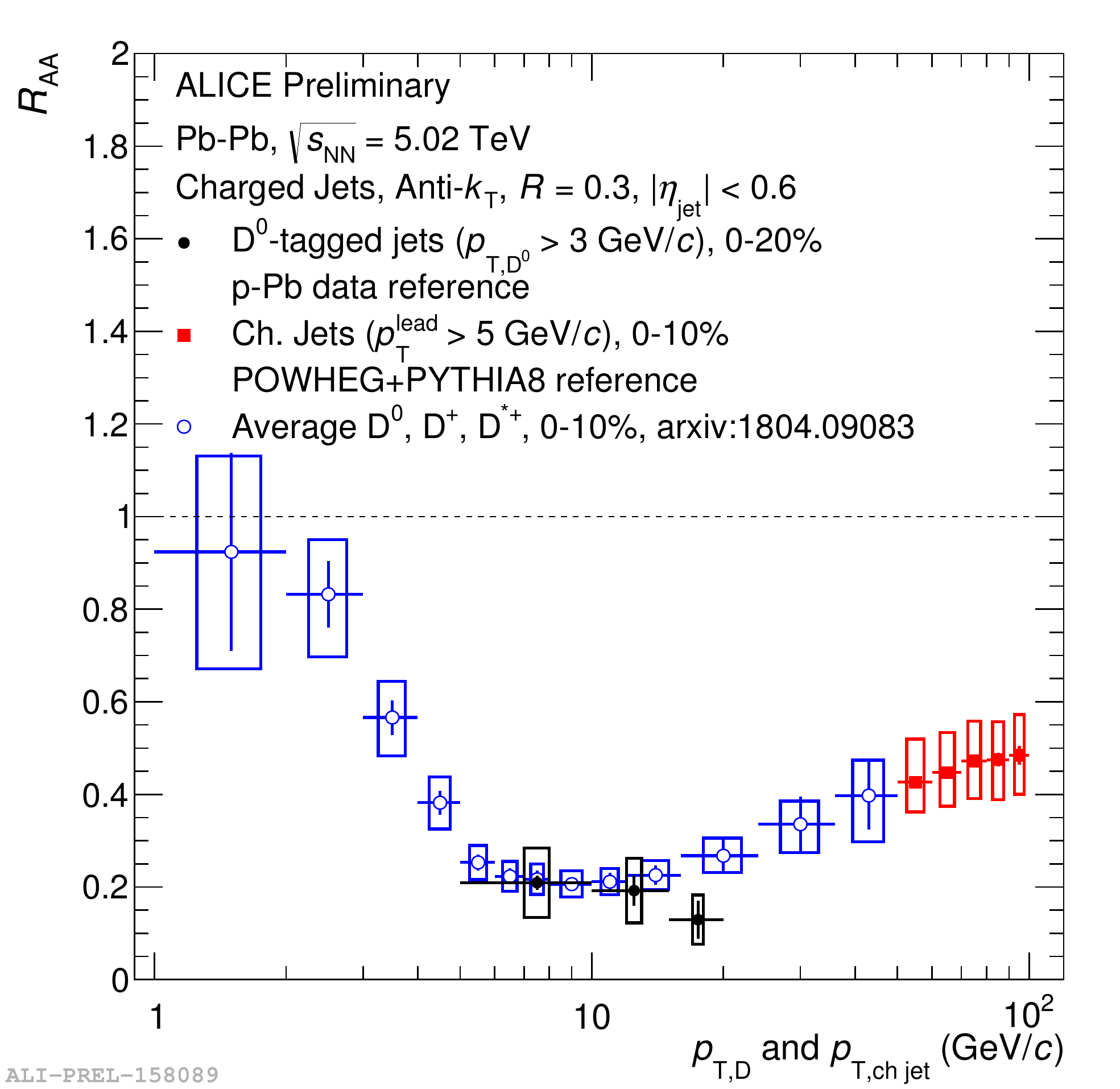}
 \includegraphics[width=0.56\textwidth]{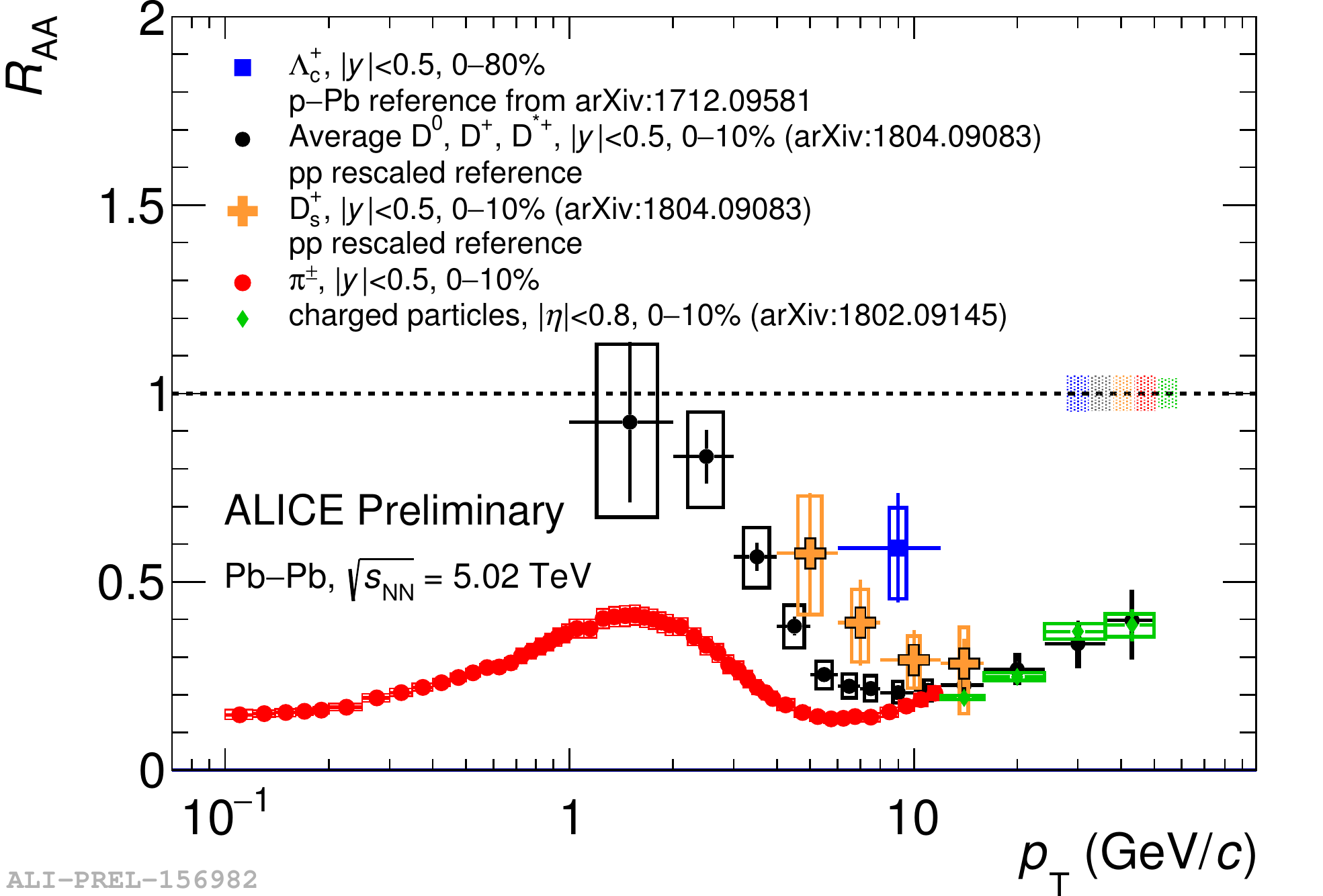}
\end{center}
\caption{Left: Comparison of the nuclear modification factor for inclusive D-mesons, jet and D-tagged jets in most central Pb–Pb collisions. Right: Comparison of the nuclear modification factor for different identified particles~\cite{HFRAA}.}
\label{fig:HFJet}
\end{figure*}

The inclusive jets are dominated by gluon- and light quark-jets. The parton energy loss is expected to be different for light and heavy flavor quarks due to their mass difference, this triggers particular interest in the separation of heavy and light flavor quark jets. The c-quark jets can be experimentally recognized by the presence of a D-hadron, creating an invariant mass peak around D meson. After the D-hadron is identified, the jet is reconstructed using the anti-$k_{\rm T}$ jet algorithm, with the D-candidate contained. With a similar correction procedure, the $D^{0}$-jet $R_{AA}$ can be calculated. Fig.~\ref{fig:HFJet} left shows the $R_{AA}$ comparison of the $D^{0}$-jets with the average D-mesons and inclusive charged jets in most central Pb-Pb collisions. While the measured $D^{0}$-jet $R_{AA}$ covers only a limited range between $5 < p_{T} < 20 $ GeV/$c$, we observe a similar suppression of $D^{0}$-tagged jets and inclusive D-meson $R_{AA}$. There is a hint of more suppression for low $p_T$ $D^{0}$-tagged jets than for inclusive jets at higher $p_T$. However, since there is no overlap range between the two measurements, one can not draw final conclusions.  

On the other hand, one indeed observed flavor dependent energy loss by studying the different particle species at lower $p_T$. Fig.~\ref{fig:HFJet} right summarizes the light and heavy flavour particles $R_{AA}$ in Pb-Pb collisions at $\sqrt{s_{NN}} = 5.02$ TeV. For $p_{T }< 10 $ GeV/$c$, there is indication of the mass dependent $R_{AA}$, namely the $R_{AA}$ of $\Lambda_{c}^{+}$ is larger than $D_{s}^{+}$, and  $D_{s}^{+}$'s $R_{AA}$ is larger than the $R_{AA}$ of $D$ and $\pi^{\pm}$. To conclude such mass hierarchy, one needs precise measurements down to very low $p_T$ with reduced systematic uncertainties. 

\section{Semi-inclusive recoil jet production}
\label{Recoils}
The measurement of reconstructed jets over a wide range in jet energy and jet resolution parameter (R) is required for the comprehensive understanding of jet quenching in heavy-ion collisions. Such measurements are challenging, however, due to the presence of complex, uncorrelated background to the jet signal, and the need to minimize biases in the selected jet population imposed by background suppression techniques.
A new approach based on the semi-inclusive distribution of charged jets recoiling from a high-$p_T$ charged hadron trigger, called "h-jet" coincidence, is deployed, where the scaling of the nuclear overlap function $T_{\rm AA}$ is not needed. This approach enables the collinear-safe measurement in heavy-ion collisions of reconstructed jets with low infrared cutoff over a wide range of jet energy and R. 
Moreover, the uncorrelated background to the recoil jet signal is corrected solely at the level of ensemble-averaged distributions, without event-by-event discrimination of jet signal from background, using a technique that exploits the phenomenology of jet production in QCD. 
$\Delta_{recoil}$ is the difference between two semi-inclusive recoil jet distributions for the Signal and Reference triggered track (TT) classes, is defined as~\cite{Recoil}:
\begin{equation}
\Delta_{\rm recoil}=\frac{1}{N_{\rm trig}^{AA}}\frac{{\rm d}^{2}N^{\rm AA}_{\rm jet}}{{\rm d} p_{\rm T, jet}^{ch}{\rm d} \eta_{\rm jet}}|_{p_{\rm T, trig}\in TT_{\rm Sig}}-c_{\rm Ref}\cdot\frac{1}{N_{\rm trig}^{AA}}\frac{{\rm d}^{2}N^{\rm AA}_{\rm jet}}{{\rm d} p_{\rm T, jet}^{ch}{\rm d} \eta_{\rm jet}}|_{p_{\rm T, trig}\in TT_{\rm Ref}},
\end{equation}
The scale factor $c_{\rm Ref}$, which is within a few percent of unity, arises because the higher TT class has a larger rate of true coincident recoil jets, and the integrals of the distributions are largely uncorrelated with TT class.

The $\Delta_{recoil}$ observable suppresses the uncorrelated jet yield in a purely data-driven way. It is directly comparable to theoretical calculations, without the need to model the heavy-ion collision background, due to utilization of a hadron trigger, the semi- inclusive nature of the observables, and the background suppression technique. The only non-perturbative component required to calculate the hard-process bias is the inclusive charged hadron fragmentation function (in-vacuum or quenched) for the trigger hadron.

\begin{figure*}[htbp]
 \begin{center}
\includegraphics[width=0.46\textwidth]{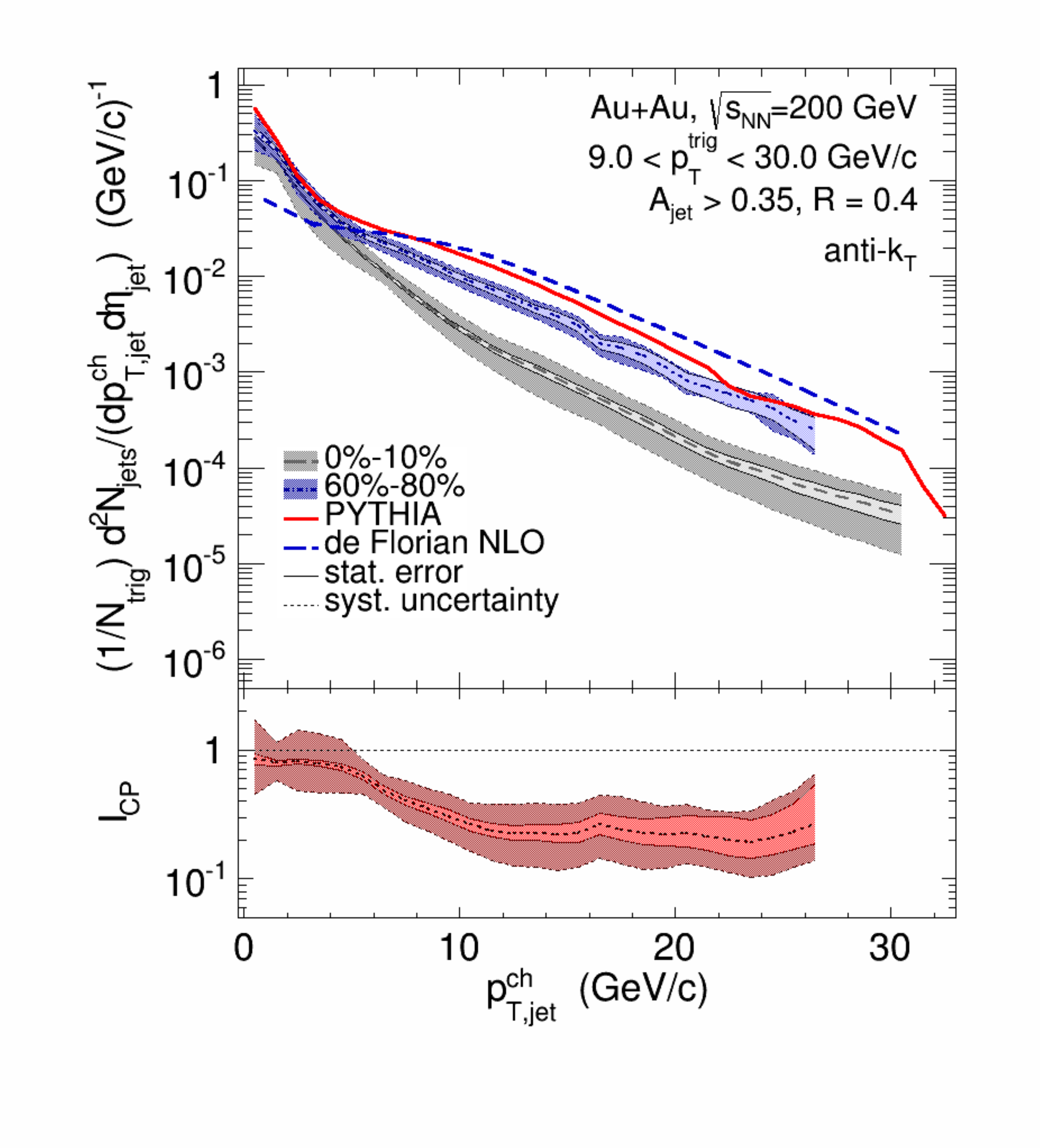}
\includegraphics[width=0.53\textwidth]{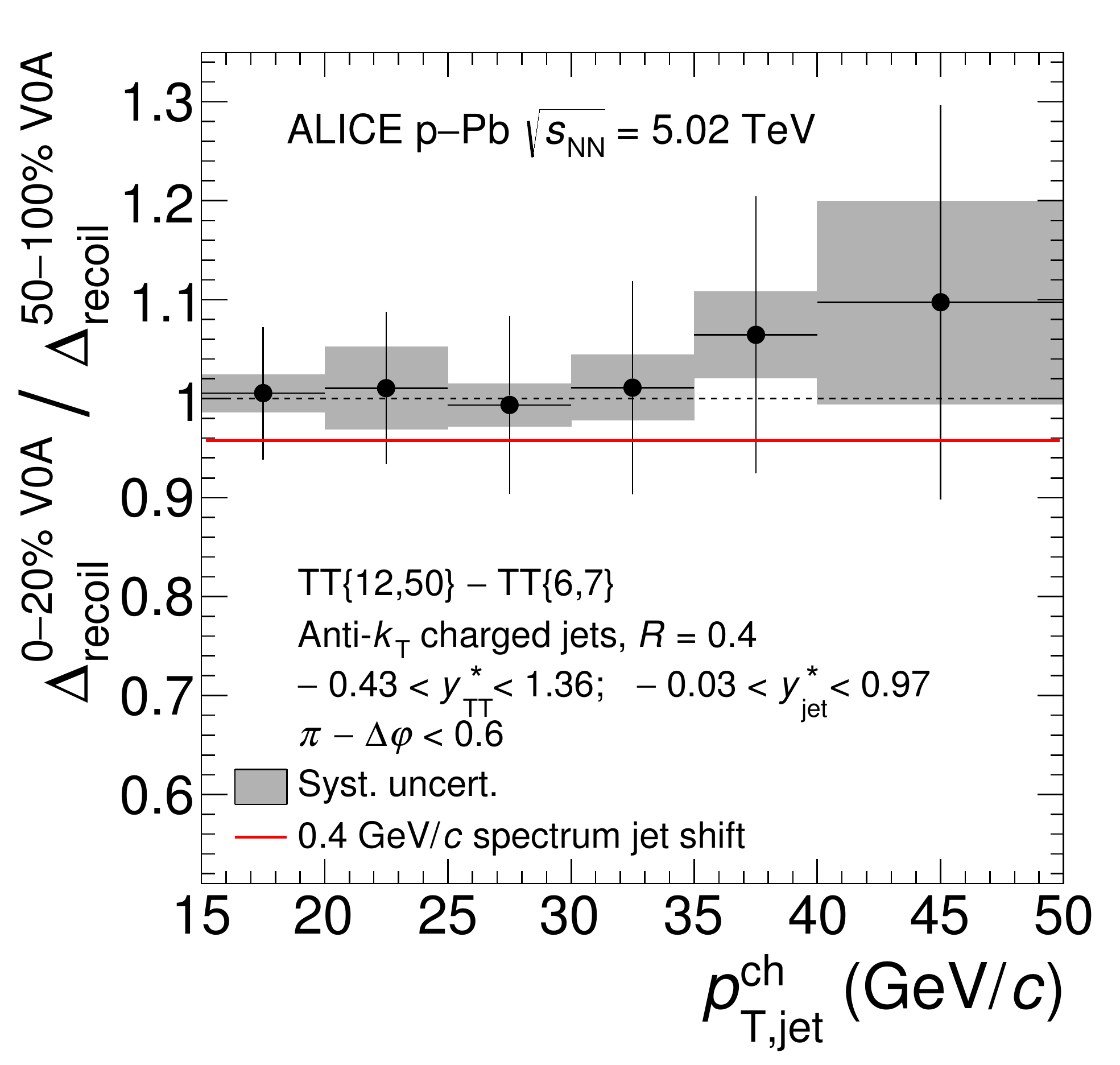}
\end{center}
\caption{ Left: The semi-inclusive recoil jet distribution in different centrality bins and compared to different model comparisons in Au-Au collisions at $\sqrt{s_{NN}} = 200$ GeV, the ratio of central to peripheral yield is shown on the bottom~\cite{STARRecoil}. Right: the ratio of $\Delta_{recoil}$ in central to peripheral p-Pb collisions at $\sqrt{s_{NN}} = 5.02$ TeV~\cite{recoilALICE}.}
\label{fig:RecoilJet}
\end{figure*}

Fig.~\ref{fig:RecoilJet} show the semi-inclusive recoil jet distribution and its ratio in different centrality classes measured by STAR in Au-Au collisions at $\sqrt{s_{NN}} = 200$ GeV and by ALICE in p-Pb collisions at $\sqrt{s_{NN}} = 5.02$ TeV with the jet resolution parameter $R=0.4$.  The automatic background removal allows for larger radii and lower jet $p_T$ reach. The semi-inclusive recoil jet yield is suppressed in central collisions compared to peripheral collisions (left)~\cite{STARRecoil}. While such $\Delta_{recoil}$ ratio in different centralities indicate more modifications in p-Pb collisions~\cite{recoilALICE}.

\section{Jet Peak Broadening}
\label{dihadron}
The measurements using two-particle angular correlations between trigger (high-$p_T$) particles and associated particles have been extensively used to search for remnants of the radiated energy and the medium response to the high-$p_T$ parton when jet reconstruction is challenging, i.e. where standard jet finding algorithms cannot be invoked event by event over the fluctuating background, especially for low $p_T$ jets. 
The presence of a "trigger" particle, having $p_T$ greater than some selected value, serves as part of the selection criteria to analyze the event for a hard scattering. By varying the transverse momentum for trigger ($p_{\rm T,trig}$) particles one can probe different momentum scales to study the interplay of soft and hard processes. 
Dihadron angular correlations represent a powerful complementary tool to study jet modifications on a statistical basis in an energy region where jets cannot be reconstructed. Such studies involve measuring the distributions of the relative azimuthal angle $\delta \phi$ and pseudo-rapidity $\delta \eta$ between particle pairs consisting of a trigger particle in a certain transverse momentum $p_{\rm T,trig}$ interval and an associated particle in a $p_{\rm T,assoc}$ interval. In these correlations, jet production manifests itself as a peak centered around $\delta \phi = 0$ and  $\delta \eta = 0 $ (near-side peak) and a structure elongated in $\delta \eta$ at $\delta \phi = \pi$ (the away side or recoil region). 

\begin{figure*}[htbp]
 \begin{center}
 \includegraphics[width=1.0\textwidth]{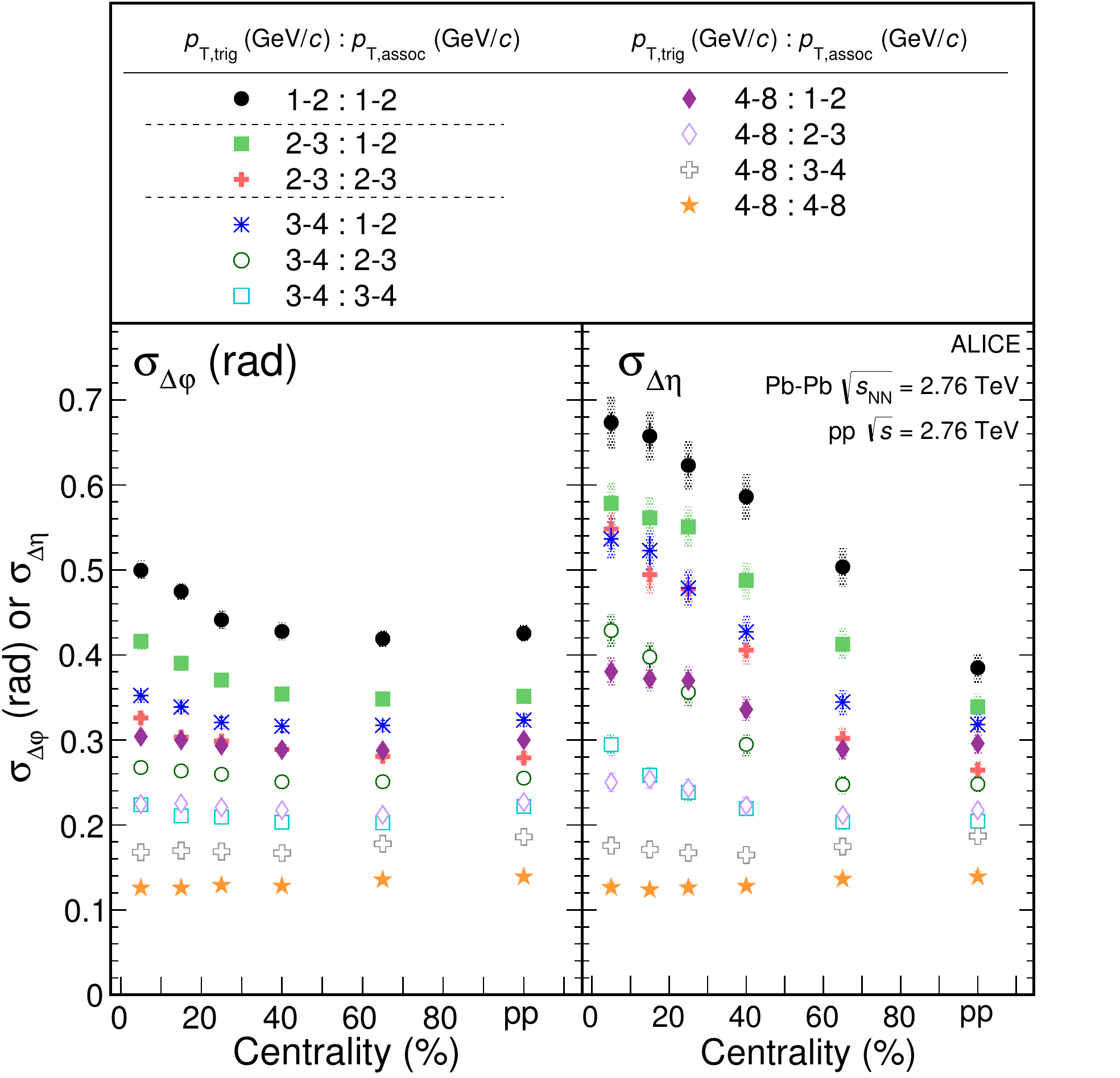}
\end{center}
\caption{Shape parameters $\sigma_{\delta \phi}$ (left panel) and $\sigma_{\delta \eta}$ (right panel) as a function of centrality in different $p_T$ ranges~\cite{ALICEJetPeak}.}
\label{fig:NearSidePeak}
\end{figure*}

\begin{figure*}[htbp]
 \begin{center}
\includegraphics[width=0.47\textwidth]{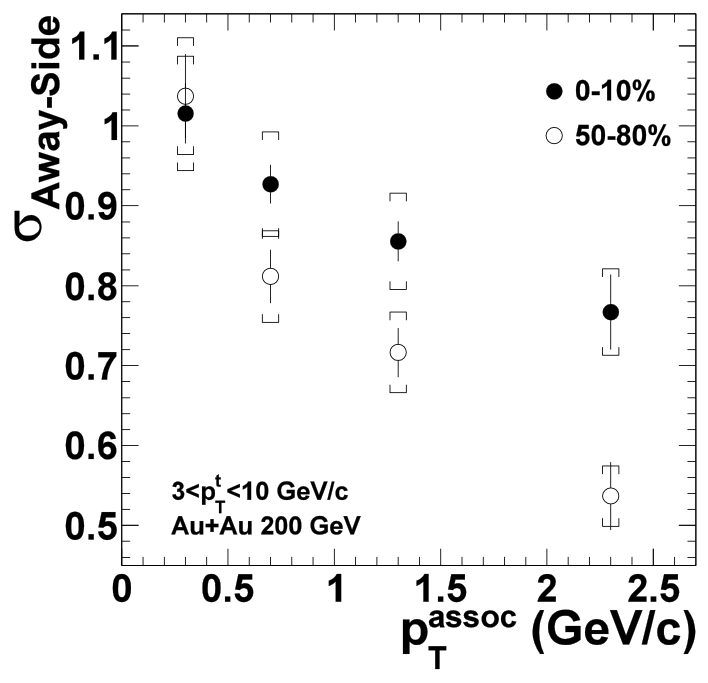}
\includegraphics[width=0.51\textwidth]{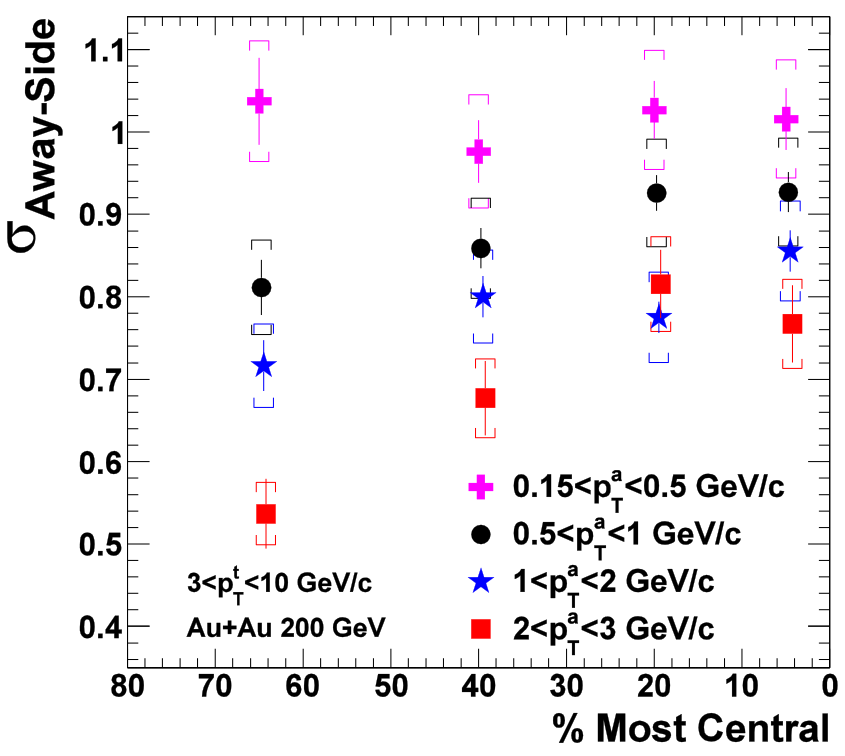}
\end{center}
\caption{Left: Shape parameters $\sigma_{\delta \phi}$ from away-side jet peak as a function of $p_{\rm T,assoc}$ in most central and most peripheral Au-Au collisions at $\sqrt{s_{NN}} = 200$ GeV. Right: Shape parameters $\sigma_{\delta \phi}$ from away-side jet peak as a function of centraility in different $p_{\rm T,assoc}$ intervals.}
\label{fig:jetBroad}
\end{figure*}

In order to characterize the near-side peak shape, a simultaneous fit of the peak, the combinatorial background, and the long-range correlation background stemming from collective effects is performed as as a function of $\delta \phi$ and $\delta \eta$. The extracted shape parameters $\sigma_{\delta \phi}$ and $\sigma_{\delta \eta}$ are presented in Fig.~\ref{fig:NearSidePeak}. The peak width increases towards central events, which is most pronounced in the lowest $p_T$ bin. In the higher $p_T$ bins no significant width increase can be observed. In the $\delta \eta$ direction (right panel) a much larger broadening is found towards central collisions~\cite{ALICEJetPeak}. Similar measurements have been performed by the STAR Collaboration, and a similar observation has been obtained as shown in Fig.~\ref{fig:jetBroad}. 

\section{Tagged jet fragmentation}
\label{ff}
The study of correlations between identified particles and jets allows us to characterize the different flavor jet fragmentation process and its possible modification in a hot nuclear matter environment. The measurement of light- and heavy-flavour jets gives more direct access to the initial parton kinematics and can provide further constraints for different quark energy loss models, in particular adding information on how the radiated energy is dissipated in the medium. It can provide information on the possible flavour dependence of jet quenching in the medium. Direct photons are produced during the early stage of the collision, through leading-order pQCD processes such as quark-gluon Compton scattering ($qg \rightarrow q\gamma$) and quark-antiquark pair annihilation
($q\bar{q} \rightarrow g\gamma$ ). In these processes the transverse energy of the trigger photon approximates the initial $p_T$ of the outgoing recoil parton, before the recoiling ("away-side") parton loses energy while traversing the medium and fragments into a jet. With the direct photon tagged jet correlations, one can probe the light quark jet fragmentation properties since Compton process is dominant. While $\pi^{0}$ associated jets or dijets mainly probe the gluon jets, photon tagged jets are originating from light quarks, and heavy flavor electrons or D and B mesons tagged jets probe heavy quark jets. Using identified particle tagged jets one can then study the jet fragmentation pattern and jet structure as a function of parton flavours.

\begin{figure*}[htbp]
 \begin{center}
\includegraphics[width=0.44\textwidth]{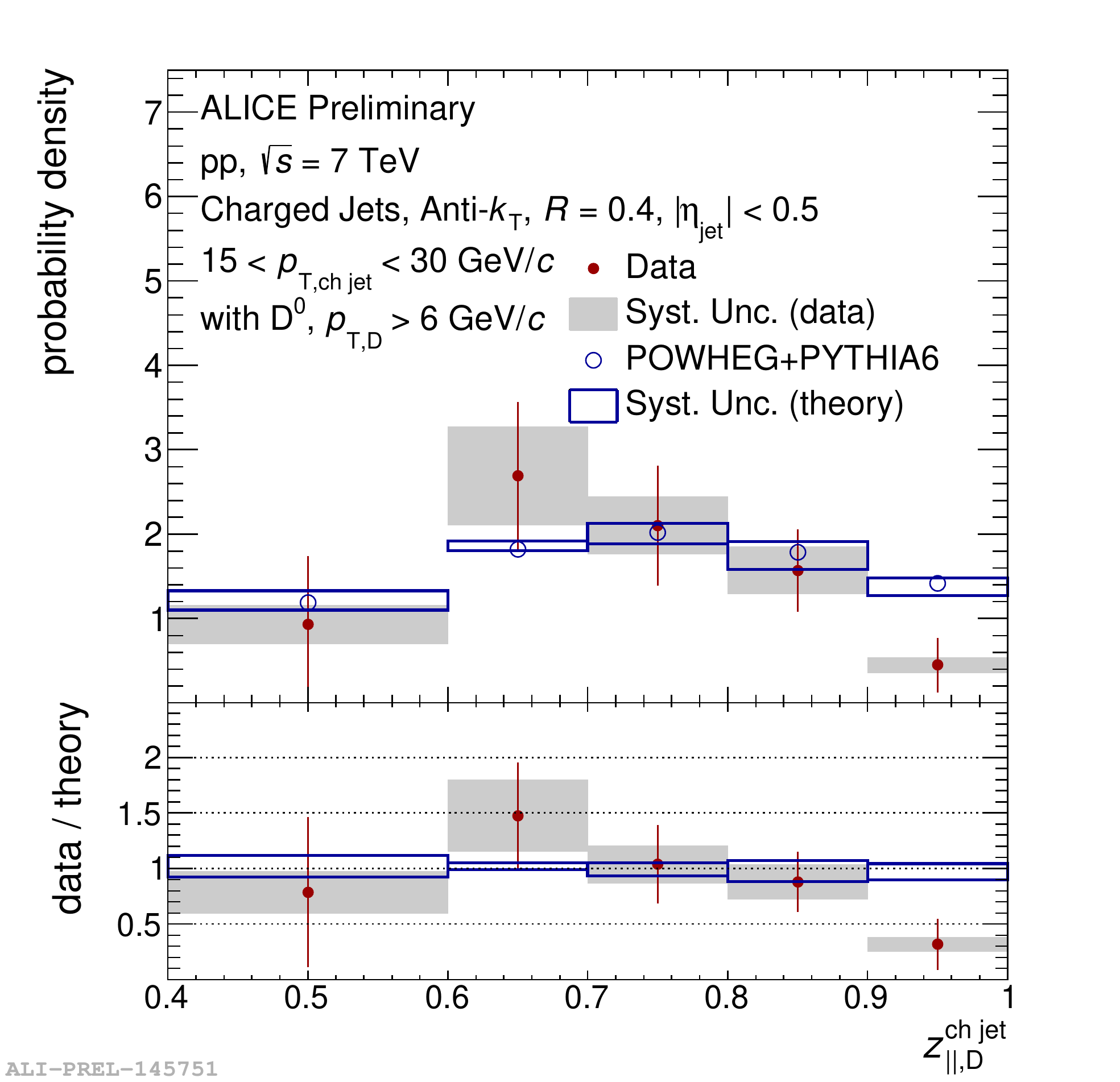}
 \includegraphics[width=0.54\textwidth]{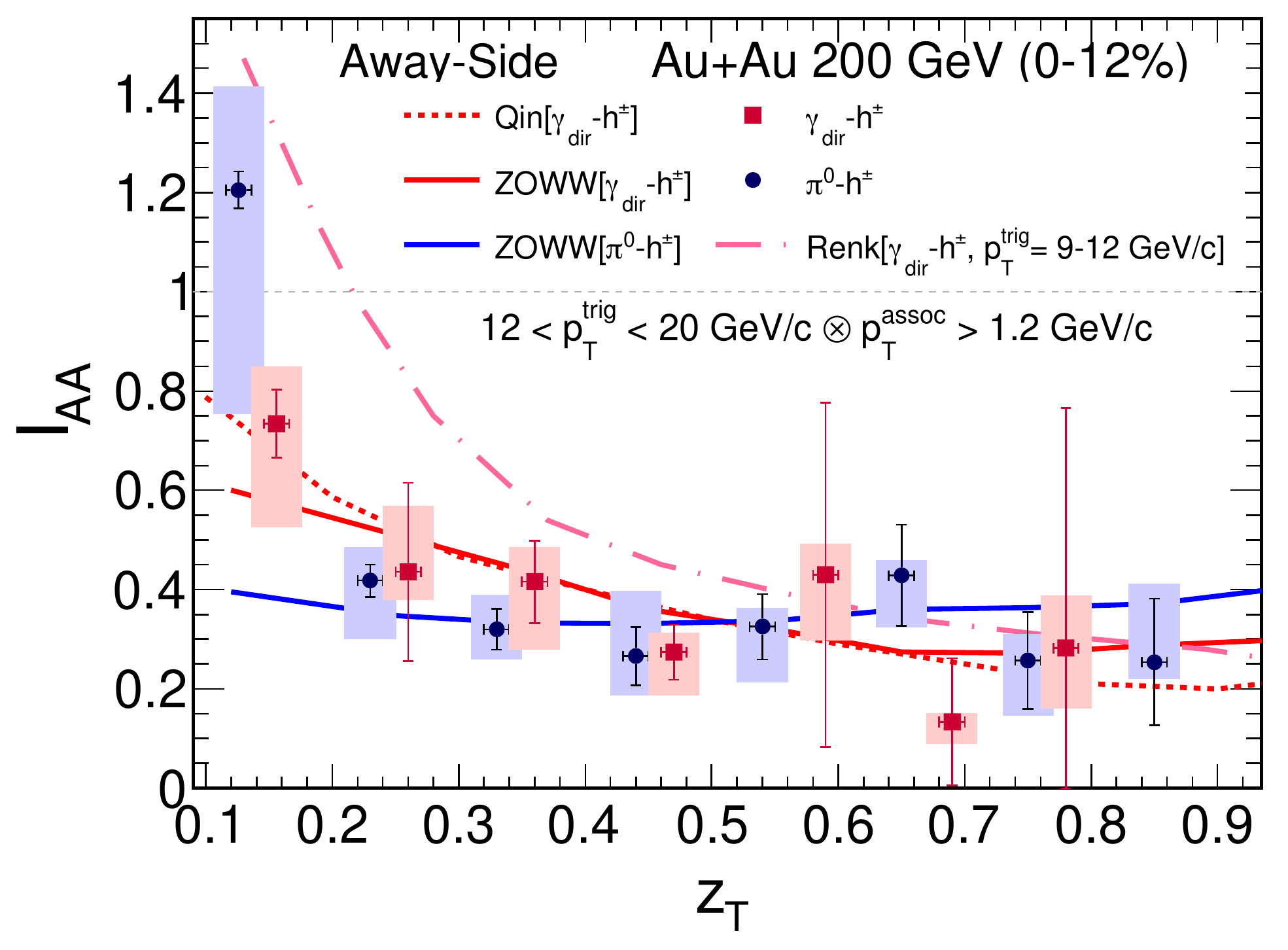}
\end{center}
\caption{Left: Probability density distribution of the jet momentum fraction, $z_{\rm ||, D}^{\rm ch, jet}$, carried by the $D^0$ mesons in the direction of the jet
axis for $D^0$-meson tagged jets in pp collisions at $\sqrt{s} = 7$ TeV for 15 $< p_{\rm T, ch jet} <$ 30 GeV/$c$~\cite{DJetFF}. Right: The $I_{AA}^{\gamma^{dir}}$ (red squares) and $I_{AA}^{\pi^0}$ (blue circles) triggers are plotted as a function of $z_T$. The points for $I_{AA}^{\gamma^{dir}}$ are shifted by +0.03 in $z_T$ for visibility~\cite{IAA}.}
\label{fig:JetFF}
\end{figure*}

Studies in pp collisions are mandatory to characterize heavy-quark production and fragmentation in vacuum, constituting the necessary reference for interpreting heavy-ion collision results. D-meson tagged charged jets are reconstructed by requiring the presence of a D meson among the jet constituents. In order to characterize the heavy quark production and fragmentation
in more detail, the distribution of the jet momentum fraction carried by $D^0$ meson in the direction of the jet axis $z_{\rm ||, D}^{\rm ch, jet} = \frac{\vec{p_{ch, jet}\dot\vec{p_{D}}}}{\vec{p_{ch, jet}\dot}\vec{p_{ch, jet}}}$ was extracted in pp collisions at $\sqrt{s} = 7$ TeV for 15 $< p_{\rm T, ch jet} <$ 30 GeV/$c$~\cite{DJetFF}. The distributions are then normalized to a probability density distribution as shown in the left panel of Fig.~\ref{fig:JetFF}. The $z_{\rm ||, D}^{\rm ch, jet}$ distribution in pp collisions are well described by the Next-to-Leading-Order pQCD calculations of POWHEG coupled with
PYTHIA6 shower MC~\cite{PYTHIA, Powheg}. 

Comparing $\gamma^{dir}$ - and $\pi^0$-triggered yields offers further opportunities to explore the geometric biases and their interplay with parton energy loss. A next-to-leading order perturbative QCD calculation~\cite{DirectPhoton} suggests that production of hadrons at different $z_T$ is also affected by different geometric biases, where $z_T = \frac{p_{T}^{assoc}}{p_{T}^{trig}}$ represents the ratio of the transverse momentum carried by a charged hadron in the recoil jet to that of the trigger particle. Using $\pi^0$ as trigger particle is an essential step for direct-photon trigger correlation since it is the dominant background for direct photon identification.  In order to quantify the medium modification for $\gamma^{dir}$ - and $\pi^0$-triggered recoil jet production as a function of $z_T$ , the ratio, $I_{AA}$ of the per-trigger conditional yields in AA to those in pp collisions is calculated. In the absence of medium modifications, $I_{AA}$ is expected to be equal to unity. Fig.~\ref{fig:JetFF} right shows the away-side
medium modification factor for $\pi^0$ triggers ($I_{AA}^{\pi^0}$) and $\gamma^{dir}$ triggers ($I_{AA}^{\gamma^{dir}}$ ), as a function of $z_T$ measured by STAR~\cite{IAA}. $I_{AA}^{\pi^0}$ and $I_{AA}^{\gamma^{dir}}$ show similar suppression within large uncertainties.

%
%
%
%

\end{document}